\documentclass[12pt,a4paper]{scrartcl}
\usepackage{enumitem}
\makeatletter
\newcommand{\mylabel}[2]{#2\def\@currentlabel{#2}\label{#1}}
\makeatother
\usepackage{braket}
\usepackage{amsfonts,amssymb}
\usepackage{color}
\usepackage{amsthm, amssymb, latexsym, amsmath,  
}
\usepackage[margin=1.2 in]{geometry}
\usepackage[ngerman,english]{babel}
\usepackage[utf8x]{inputenc}
\usepackage[OT2,T1]{fontenc}

\newtheorem{theorem}{Theorem}[section]

\newtheorem{corollary}[theorem]{Corollary}

\newtheorem{remark}{Remark}

\newcommand{\DETAILS}[1]{}

\newcommand{\R}{\mathbb{R}}

\newcommand{\N}{\mathbb{N}}

\newcommand{\cA}{{\cal{A}}}

\newcommand{\cH}{{\cal{H}}}

\newcommand{\cO}{{\cal{O}}}

\newcommand{\cE}{\mathcal{E}}

\newcommand{\vphi}{{\varphi}}
\newcommand{\e}{{\mathrm{e}}}

\newcommand{\scp}[2]{\langle #1\text{,}#2\rangle}
\newcommand{\tr}[1]{\mathrm{Tr}#1}

\begin{document}

\title{A simple proof of convergence to the Hartree dynamics  in Sobolev trace norms}

\author{Ioannis Anapolitanos\thanks{Dept.~of Math.,
		Karlsruhe Institute of Technology, Karlsruhe, Germany,}, Michael Hott\thanks{Dept.~of Math., Karlsruhe Institute of Technology, Karlsruhe, Germany, and Dept.~of Math.
		University of Texas at Austin, Austin/TX, USA.}
	}
\bigskip

\bigskip

\bigskip

\bigskip
\bigskip

\maketitle

\begin{abstract}
	The derivation of the Hartree equation from many-body systems of Bosons in the mean field limit has been very intensively studied in the last couple of years. However, very few results exist showing convergence of the $k$-th marginal of the $N$-body density matrix to the projection to the $k$-fold tensor product of the solution of the Hartree equation in stronger trace norms like the energy trace norm, see \cite{MS}, \cite{Lu}. This issue is from a physical view point very important. The reason is that one can then approximate expectation values of certain observables of the $N$-body system  by means of the Hartree equation, with relaxation of the very restrictive assumption that the observables are bounded operators. Here we consider the non-relativistic case. We prove, assuming only $H^1$-regularity of the initial data, convergence in the energy trace norm without rates, and convergence in any other weaker Sobolev trace norm with rates. Our proof is simple and uses the functional $a_N$ introduced by Pickl in \cite{Pi}. 
\end{abstract}

\section{Introduction}

In this work, we consider a system of $N$ Bosons in $\R^3$ described by the Hamiltonian
\begin{equation}\label{totham}
H_N\;:=\;\sum_{i=1}^N T_i +\frac{1}{N-1}\sum_{i<j}v(x_i-x_j),
\end{equation}
where 
\begin{equation}\label{def:v}
v(x)=\lambda \frac{\e^{-\mu|x|}}{|x|} \text{ with } \lambda\in\R \text{ and } \mu \geq0. 
\end{equation}
We focus on the case $T=-\Delta$.
The Hamiltonian $H_N$ is considered on the Hilbert space
\begin{equation*}
\cH_N\;:=\;L^2(\R^3)^{\otimes_S N},
\end{equation*}
where $\otimes_S$ denotes the symmetric tensor product. 
We are interested in  the dynamics of  a Bose-Einstein Condensate. More precisely we assume that there is $\vphi_0\in L^2(\R^3)$ with $\|\vphi_0\|_2=1$ such that the many-body wave function of the system is given by the solution of 
 \begin{equation}\label{nbodyeq}
\begin{cases}
i\partial_t\Psi_{N,t}\;&=\;H_N\Psi_{N,t},\\
\Psi_{N,0}\;&=\;\vphi_0^{\otimes N}.
\end{cases}
\end{equation}
 An effective equation for \eqref{nbodyeq} is the \textit{Hartree equation} given by
\begin{equation}\label{hartree}
\begin{cases}
i\partial_t\vphi_t\;&=\;T \vphi_t+v*|\vphi_t|^2\vphi_t\\
\vphi_{t=0}\;&=\;\vphi_0
\end{cases}
\end{equation}
on $L^2(\R^3)$. For $T=-\Delta$ the global well-posedness of \eqref{hartree} in $H^1(\R^3)$ is known. It follows by standard fixed point arguments and the conservation of $\|\vphi_t\|_{L^2}$ and  of the energy functional $\cE(\vphi_t)$ given by  $$\mathcal{E}(\vphi):=\int | \nabla \vphi(x)|^2dx + \frac{1}{2} \int |\vphi(x)|^2 v(x-y) |\vphi(y)|^2 dx dy.$$
In the proof of the well-posedness Hardy's inequality in $\R^3$, namely $\frac{1}{|x|^2} \leq -4 \Delta$, is also used. With the help of it one can bound the  $H^1(\R^3)$ norm of $\vphi_t$ in terms of $\mathcal{E}(\vphi_t)=\mathcal{E}(\vphi_0)$ and it follows that there exists $C=C(\vphi_0)$ such that
\begin{equation}\label{est:H1}
\|\vphi_t\|_{H^1(\R^3)}\leq C, \quad \forall t>0.
\end{equation}
  A typical result showing that \eqref{hartree} is indeed an effective equation for \eqref{nbodyeq} is the following: If $\cA$ is a bounded and self-adjoint operator on $L^2(\R^{3})^{\otimes_Sk}$ with $k\in\N$, then there exist constants $C_{k,t,\cA,\vphi_0}>0$ and $0<\alpha\leq1$ such that the following inequality holds for all $N \geq k$:
\begin{equation}\label{aeff}
\left|\scp{\Psi_{N,t}}{\cA\otimes I_{N-k}\Psi_{N,t}}-\scp{\vphi_t^{\otimes k}}{\cA\vphi_t^{\otimes k}}\right|\leq \frac{C_{k,t,\cA,\vphi_0}}{N^\alpha}.
\end{equation}
The physical interpretation of \eqref{aeff} is that the expectation of the observable $\cA$ in the $N$-body system can be approximated by means of the Hartree equation, which is much easier to handle numerically because it is a one-body equation. Such results are given e.g. in \cite{CLS}, \cite{ES}, \cite{KP}, \cite{RS}, \cite{Le}. For other related results see \cite{AFP}, \cite{AN}, \cite{ElS}, \cite{GV}, \cite{He},  \cite{Sp} and references therein, to name a few. There are very few results showing \eqref{aeff} in the case that $\cA$ is an unbounded observable, e.g. the potential energy or the momentum and this is the goal of this work.  Michelangeli and Schlein were the first to discuss this case in \cite{MS}. They obtained in the semi-relativistic setting ($T=\sqrt{1-\Delta}$) an estimate of the form
\begin{equation}\label{unbddest}
\left|\scp{\tilde{\Psi}_{N,t}}{\cA\otimes I_{N-k}\tilde{\Psi}_{N,t}}-\scp{\vphi_t^{\otimes k}}{\cA\vphi_t^{\otimes k}}\right|$$$$\leq \frac{C_{k,t,\vphi_0} \|(1+\sum_{i=1}^k T_i)^{-\frac12}\cA(1+\sum_{i=1}^k T_i)^{-\frac12}\|}{N^\alpha}
\end{equation}
for some $0<\alpha\leq 1$ to show  dynamical collapse of boson stars, where $\tilde{\Psi}_N$ is, in a certain sense, an approximate solution of \eqref{nbodyeq} \footnotemark{\footnotetext{$\tilde{\Psi}_N$ is the evolution due to a regularized Hamiltonian because $H_N$ is not always a self-adjoint operator in this setting. We refer to \cite{MS} for details.}.
 Adopting the methods of \cite{MS}, Lührmann proved in \cite{Lu} an analogous result for the magnetic Hartree equation, where $T$ is the magnetic Laplacian. Michelangeli and Schlein require $H^2$-regularity of the initial data and Lührmann requires $H_A^3$-regularity, where $H_A^3$ denotes the magnetic Sobolev space of order 3. Before we formulate our first main theorem, we quickly introduce some useful notation associated with the solutions of \eqref{nbodyeq} and \eqref{hartree}: 
\begin{align}
\gamma_{N,t}^{(k)}\;&:=\;\tr_{k+1,...,N}|\Psi_{N,t} \rangle \langle \Psi_{N,t}|,\nonumber\\
P^{(k)}_t\;&:=\;|\vphi_t^{\otimes k} \rangle \langle \vphi_t^{\otimes k}|,\nonumber\\
S_{k}\;&:=\;\sum_{i=1}^k (1 - \Delta_{x_i} )\label{eq:skmag},
\end{align}
and as in \cite{Pi}, 
\begin{equation}\label{def:aNt}
a_{N,t}:=\langle \Psi_{N,t}, q_1^{\vphi_t} \Psi_{N,t} \rangle,
\end{equation}
 where $q_1^{\vphi_t}=1-p_1^{\vphi_t}$, with $p_1^{\vphi_t}=(\ket{\vphi_t}\bra{\vphi_t})_1$.
We are now ready to state our first main result:

\begin{theorem}\label{thm:main}
	Assume that $T=-\Delta$ and that $\vphi_0 \in H^1$. 
	\begin{itemize}
		\item[\bf (i)] For any $\theta \in (0,1)$ there exists a constant $C>0$ such that for any $k\in\N$, $N\geq k$ and any $t>0$ we have
		\begin{equation}\label{es:preexpli}
		\tr\left|S_{k}^{\frac \theta 2} (\gamma_{N,t}^{(k)}-P^{(k)}_t) S_{k}^{\frac \theta 2}\right| \leq C k(a_{N,t}^{\min(\frac{1}{2}, 1-\theta)} +  \|\gamma_{N,t}^{(k)}-P^{(k)}_t\|_{HS}^{1-\theta}).
		\end{equation}
			\item[\bf (ii)] For all $k \in \N$ and $t>0$ we have
			\begin{equation}\label{traceenergyconv}
			\lim_{N \rightarrow \infty} \tr\left|S_{k}^{\frac12} (\gamma_{N,t}^{(k)}-P^{(k)}_t) S_{k}^{\frac12}\right|\;=\;0. 
			\end{equation}
	\end{itemize}
\end{theorem}
$\tr{\left|S_{k}^{\frac12}(\cdot)S_{k}^{\frac12}\right|}$ is called the energy trace norm of operators acting on $L^2(R^3)^{\otimes_S k}$ and it is associated with the kinetic energy of the systems and the Sobolev space $H^1$. The left hand side of \eqref{es:preexpli} is a Sobolev trace norm associated with the Sobolev space $H^\theta$.
Theorem \ref{thm:main} has several corollaries which can be obtained if one includes previous results estimating $a_{N,t}$ and  $\|\gamma_{N,t}^{(k)}-P^{(k)}_t\|_{HS}^{1-\theta}$. Among others, such results are the following:
\vspace{1ex}
\begin{theorem}[Pickl \cite{Pi}, Knowles, Pickl \cite{KP}]
	Under the assumptions of Theorem \ref{thm:main} there exist constants $C,D\in\R$ independent of $N,t$ such that for any $N \in \N$ and $t >0$ we have  $a_{N,t} \leq \frac{C e^{Dt}}{N}$. 
\end{theorem}
\vspace{1ex}
\begin{theorem}[Chen, Lee, Schlein \cite{CLS}]\label{thm:CLS}
	We assume again the conditions of Theorem \ref{thm:main}. Then for any $k \in \N$ there exist $C_k,D_k \in \R$ such that for any $N\in\N$ with $N \geq k$ and any $t>0$ we have  $ \|\gamma_{N,t}^{(k)}-P^{(k)}_t\|_{HS} \leq \frac{C_k e^{D_kt}}{N}$. 
\end{theorem}
\vspace{1ex}
Using Theorem \ref{thm:main} together with these theorems and the fact that the dual space of trace class operators are the bounded operators, we obtain the following corollary:
\vspace{1ex}
\begin{corollary}\label{cor:application}
	Assume $T=-\Delta$ and $\vphi_0 \in H^1$. 
	 Let $k \in \N$ and $\cA$ be a self-adjoint operator acting on $L^2(\R^3)^{\otimes_S k}$. Assume that there exists  $\theta \in [0,1]$ such that 
	$ S_{k}^{-\frac \theta 2} \cA S_k^{-\frac{\theta}{2}}$ can be extended to a bounded operator on $L^2(\R^3)^{\otimes_S k}$ with operator norm $\|S_{k}^{-\frac \theta 2} \cA S_k^{-\frac{\theta}{2}}\|$.
	\begin{itemize}
	\item[\bf (i)] If $\theta<1$, there exist $C_k, D_k > 0$ independent of $N,t$ such that for any $N\in\N$ with $N \geq k$ and any $t>0$ we have
	\begin{equation}\label{es:explicit}
	\left|\scp{\Psi_{N,t}}{\cA\otimes I_{N-k} \Psi_{N,t}}-\scp{\vphi_t^{\otimes k}}{\cA\vphi_t^{\otimes k}}\right| \leq \frac{C_k e^{D_k t}}{N^{\min(\frac{1}{2}, 1-\theta)}} \| S_{k}^{-\frac \theta 2} \cA S_k^{-\frac{\theta}{2}}\|.
	\end{equation}
		\item[\bf (ii)] If $\theta=1$, then
		\begin{equation}
		\lim_{N \rightarrow \infty}\left|\scp{\Psi_{N,t}}{\cA\otimes I_{N-k} \Psi_{N,t}}-\scp{\vphi_t^{\otimes k}}{\cA\vphi_t^{\otimes k}}\right| = 0 \quad \forall t>0.
		\end{equation}
		\end{itemize}
\end{corollary}

This corollary shows the physical importance of Theorem \ref{thm:main}. Previous theorems estimating the left hand side of $\eqref{es:explicit}$ only for bounded $\cA$ can be combined with our theorem to estimate the left hand side of $\eqref{es:explicit}$ for a much larger class of observables $\cA$. As a consequence, expectations of various physical observables like the momentum and the kinetic and potential energy of the $N$-body system can be approximated by means of the Hartree equation.

\begin{remark}
	To our best of knowledge Theorem \ref{thm:main} is the first result of convergence in Sobolev trace norms without assuming  regularity of the initial data which is stronger than the natural $H^1$ regularity. Moreover, even with stronger regularity assumptions of the initial data,  estimate \eqref{es:explicit} does not immediately follow from the methods in \cite{MS} and \cite{Lu}. The reason is that if for example $\vphi_0 \in H^3$ then there is only a superexponential bound in time for the growth of $\|\vphi_t\|_{H^3}$, see Lemma 3 in  \cite{Len}. Such bound would lead to a time dependence worse than in the right hand side of \eqref{es:explicit} even if the $N$ dependence were better. Moreover, the method in the proof of  Theorem \ref{thm:main} is much simpler than the methods used  in \cite{MS} and \cite{Lu}. However, we also want to emphasize the fact that unlike in \cite{Lu} and \cite{MS} we were not able to find with our methods any rates of convergence in the case of the energy trace norm even assuming additional regularity of the initial data. We were also not able to show that our method can be applied to the case considered by \cite{MS}, where the potential energy of the $N$-body system can be attractive and does not have to be $\sum_{j=1}^N(\sqrt{1-\Delta_{x_j}})$-form bounded with relative bound less than one. 
\end{remark}

\begin{remark}
	There are several  generalizations of Theorem \ref{thm:main} some of them quite obvious. A slight modification of the proof shows that the theorem remains true if the assumption $\Psi_{N,0}=\vphi_0^{\otimes N}$  is replaced by the weaker condition $$\lim_{N \to \infty}\tr|\gamma_{N,0}^{(1)}-P^{(1)}_0| = 0 \quad \text{ and } \quad \lim_{N \to \infty}\frac{1}{N} \langle \Psi_{N,0}, H_N \Psi_{N,0} \rangle=\mathcal{E}(\vphi_0).$$
	For part (i) to be true we can even relax the second part of this weaker condition by only assuming that the sequence $\frac{1}{N} \langle \Psi_{N,0}, H_N \Psi_{N,0} \rangle$ is bounded.  
	Furthermore, Theorem \ref{thm:main} remains true if one uses any scaling $\cO(N^{-1})$ instead of $(N-1)^{-1}$ in front of the Coulomb interaction in \eqref{totham}. Moreover the potential $v$ does not have to be the Coulomb potential, an assumption of the form $v^2 \leq C (1-\Delta)$ for some $C>0$ would be enough.
	Replacing the Laplacian by the magnetic Laplacian is also possible under certain assumptions on the magnetic potential, this will be discussed in detail in \cite{AHH}. Under some further assumptions, we can consider also the semi-relativistic case $T=\sqrt{1-\Delta}$ and obtain an analogous result to part (i) of Theorem \ref{thm:main}, which will also be discussed in \cite{AHH}. Estimate \eqref{es:preexpli} is a consequence of Theorem \ref{thm:main1} below, which is a general interpolation argument between different Sobolev trace norms and we find it of interest on its own, because it could be a helpful tool to upgrade convergence in trace norm to convergence in Sobolev trace norms in other situations.
\end{remark}

\par The rest of the paper is organized as follows: In Section \ref{sec:thmspfs} we start with the proof of  part (i) of Theorem \ref{thm:main}. To this end we first state and prove Theorem \ref{thm:main1}, our second main theorem, and part (i) of Theorem \ref{thm:main} follows as a corollary. We then prove part (ii) of Theorem \ref{thm:main}.

\noindent {\bf Acknowledgments.} We are grateful to Dirk Hundertmark for numerous stimulating discussions and for the suggestion to work with the Pickl functional $a_N$ in the mean-field context. We gratefully acknowledge financial support by the Deutsche Forschungsgemeinschaft (DFG) through CRC 1173.

\section{Proof of Theorem \ref{thm:main}}\label{sec:thmspfs}

\subsection{Proof of part (i) of Theorem \ref{thm:main} }

Instead of part (i) of Theorem \ref{thm:main}, we first prove a more general result which can be used for interpolation between Sobolev trace norms. 
	 Let $s>0$, $\Psi \in \cH_N \cap H^s(\mathbb{R}^{3N})$, $\vphi \in H^s(\mathbb{R}^{3})$ with $\|\Psi\|_{L^2}=\|\vphi\|_{L^2}=1$. For $k \in \{1, \ldots, N\}$ we define $$\gamma_N^{(k)}=\tr_{k+1,...,N}|\Psi \rangle \langle \Psi|,$$ 
$P^{(k)}:=|\vphi^{\otimes k} \rangle \langle \vphi^{\otimes k}|$  and, as in \cite{Pi}, 
\begin{equation}\label{def:aN}
a_N:=\langle \Psi, q_1 \Psi \rangle = \|q_1 \Psi\|^2,
\end{equation}
 where we write $q_1:=q_1^\vphi=1-p_1^\vphi$, with $p_1^\vphi=(\ket{\vphi}\bra{\vphi})_1$. We obviously have
 \begin{equation}\label{pickltrace}
 a_N\;=\;\tr{\left(p_1(p_1-\gamma_N^{(1)})\right)}\;\leq\;\tr\left|\gamma_N^{(1)}-p_1\right|.
 \end{equation}
Let for $r \in \R$ 
\begin{equation}\label{def:skr}
S_{k,r}:=\sum_{i=1}^k (1-\Delta_{x_i})^{r}
\end{equation}
and we denote the Hilbert-Schmidt norm of an operator acting on $L^2$ with $\|.\|_{HS}$. Now we are ready to formulate our second main result. 
\\[1ex]
\begin{theorem}\label{thm:main1}
	For any $\theta \in [0,1)$ we have the estimate
	\begin{equation*}
	\tr\left|S_{k,\theta s}^{\frac12} (\gamma_N^{(k)}-P^{(k)}) S_{k,\theta s}^{\frac12}\right|\;\leq\; kC_{\Psi,\vphi,\theta,s} ( a_N^{\min(\frac{1}{2},1-\theta)} + \|\gamma_N^{(k)}-P^{(k)}\|_{HS}^{1-\theta}),
	\end{equation*}
where $C_{\Psi,\vphi,\theta,s}:=	2 \max\left(\|S_{1,s}^\frac{1}{2}\Psi\|_2 +\||S_{1,s}^\frac{1}{2}\vphi\|_2,(\||S_{1,s}^\frac{1}{2}\Psi\|_2+\||S_{1,s}^\frac{1}{2}\vphi\|_2)^{2 \theta} \right)$.
\end{theorem}
\begin{remark}\label{rem:intuition}
	The case $s=1$, which we are going to use to prove part (i) of Theorem \ref{thm:main}, has a simple intuition. We observe that $C_{\Psi,\vphi,\theta,1}$ is associated with the kinetic energy per particle of $\Psi$ and the kinetic energy of $\vphi$. If one considers dynamics and manages to control these kinetic energies  uniformly in $N$, then convergence in trace norm can be automatically upgraded to convergence of Sobolev trace norms up to but not including the energy trace norm. However, we state Theorem \ref{thm:main1} for general $s>0$ for several reasons. First, as we will discuss in \cite{AHH} the case $s=\frac{1}{2}$ has applications in the semi-relativistic case. Second, it is unclear to us whether one can control $C_{\Psi_{N,t},\vphi_t,\theta,s}$ uniformly in $N$ for $s>1$ in the non-relativistic case respectively for $s>\frac12$ in the semi-relativistic case, if one assumes higher regularity of the initial data. Uniform boundedness of $C_{\Psi_{N,t},\vphi_t,\theta,s}$ in $N$ would imply convergence in the energy trace norm with rates.  
\end{remark}
\vspace*{1ex}
\begin{proof}
	First of all, we abbreviate for all $r\in\R$
	$$A_{k,r}:=S_{k,r}^{\frac12} (\gamma_N^{(k)}-P^{(k)}) S_{k,r}^{\frac12}.$$
	We divide the proof into three steps. In \textit{step 1} we show
	\begin{equation}\label{est:stongtrace1}
	\tr|A_{k,\theta s}| \leq 2 \|A_{k,\theta s}\|_{HS} + \tr(A_{k,\theta s}).
	\end{equation}
	In \textit{step 2}, we show	\begin{equation}\label{est:stongtrace2}
	\|A_{k,\theta s}\|_{HS} \leq k (\|S_{1,\frac{s}2}\Psi\|_2+\|S_{1,\frac{s}2}\vphi\|_2)^{2 \theta} \|\gamma_N^{(k)}-P^{(k)}\|_{HS}^{1-\theta}.
	\end{equation}
	In the last step, we prove 
	\begin{equation}\label{est:stongtrace3}
	\tr(A_{k,\theta s})\leq k \max\left(\|S_{1,\frac{s}2}\Psi\|_2+\|S_{1,\frac{s}2}\vphi\|_2,(\|S_{1,\frac{s}2}\Psi\|_2+\|S_{1,\frac{s}2}\vphi\|_2)^{2\theta} \right) a_N^{\min(\frac{1}{2},1-\theta)}.
	\end{equation}
	
	Obviously these three steps imply Theorem \ref{thm:main1}.
	
	\underline{Step 1:} The argument to prove \eqref{est:stongtrace1} is known and was observed by Robert Seiringer as mentioned in \cite{RS}. We repeat it here for convenience of the reader. Since $P^{(k)}$ is a rank-one projection in the $k$-particle space, the variational characterization of eigenvalues implies that $A_{k,\theta s}$ has at most one negative eigenvalue. If there are no negative eigenvalues, then $\tr|A_{k,\theta s}| = \tr(A_{k,\theta s})$ so \eqref{est:stongtrace1} trivially holds. We may therefore assume that $A_{k,\theta s}$ has a negative eigenvalue $\lambda_1$ and let $(\lambda_n)_{n\geq2}$ be the sequence of its nonnegative eigenvalues counting multiplicity. It follows that
	$\tr|A_{k,\theta s}|=|\lambda_1|+\sum_{n=2}^\infty \lambda_n= 2 |\lambda_1|+ \tr(A_{k,\theta s})$. Since $|\lambda_1| \leq \|A_{k,\theta s}\|_{HS}$, the upper bound	\eqref{est:stongtrace1} follows. 
	
	\underline{Step 2:} Let $L(x,y)$ denote the integral kernel of $\gamma_N^{(k)}-P^{(k)}$, where $x=(x_1,\dots,x_k)$, $y=(y_1,\dots,y_k)$ are elements of $\R^{3k}$. Then $A_{k,\theta s}$ has an integral kernel given by
	$$\left[\sum_{i=1}^k (1-\Delta_{x_i})^{\theta s}\right]^\frac12 \left[\sum_{j=1}^k (1-\Delta_{y_j})^{\theta s}\right]^\frac12 L(x,y).$$ 
	Therefore, using that the Hilbert-Schmidt norm of an operator is equal to the $L^2$-norm of its kernel and Plancherel's Theorem, we obtain
	\begin{equation*}
	\|A_{k,\theta s}\|_{HS}^2\;=\;\int\left[\sum_{i=1}^k (1+|\xi_i|^2)^{\theta s}\right] \left[\sum_{j=1}^k (1+|\eta_j|^2)^{\theta s}\right] |\hat{L}(\xi,\eta)|^2d\xi d\eta,
	\end{equation*}
	where $\xi=(\xi_1,\ldots,\xi_k)$, $\eta=(\eta_1,\ldots,\eta_k)$
	and $\hat{L}(\xi,\eta)=\frac{1}{(2 \pi)^\frac32}\int e^{- i (x \xi+ y \eta)} L(x,y) dx dy$.
	 Since $t\mapsto t^\theta$ is concave, it follows that
	$$\sum_{i=1}^ka_i^\theta\;\leq\;k^{1-\theta}\left(\sum_{i=1}^ka_i\right)^\theta\quad\forall a_1,\ldots,a_k\geq0,$$
	and hence
	\begin{equation*}
	\|A_{k,\theta s}\|_{HS}^2\;\leq\; k^{2(1-\theta )} \int\left[\sum_{i=1}^k (1+|\xi_i|^2)^{s}\right]^\theta  \left[\sum_{j=1}^k (1+|\eta_j|^2)^{s}\right]^\theta  |\hat{L}(\xi,\eta)|^2 \mathrm{d}(\xi,\eta).
	\end{equation*}	
	Thus applying Hölder's inequality and Plancherel's Theorem again, we arrive at
	\begin{equation}\label{est:Hoelderfourier1}
	\|A_{k,\theta s}\|_{HS} \leq \|A_{k,s}\|_{HS}^{\theta}\|A_{k,0}\|_{HS}^{1-\theta}.
	\end{equation}
	Furthermore, we have
	\begin{equation}\label{eq:akshs}
	\begin{split}
	\|A_{k,s}\|_{HS}\;& \leq\; \tr|A_{k,s}| 
	\leq \tr(S_{k,s}^\frac{1}{2} \gamma_N^{(k)} S_{k,s}^\frac{1}{2})+ \tr(S_{k,s}^\frac{1}{2} P^{(k)} S_{k,s}^\frac{1}{2})\\
	&=\; \langle \Psi, S_{k,s} \Psi\rangle+\langle \vphi^{\otimes k}, S_{k,s} \vphi^{\otimes k} \rangle\\
	&=\;k (\langle \Psi, S_{1,s} \Psi\rangle+\langle \vphi, S_{1,s} \vphi \rangle)
	\end{split}
	\end{equation}
	So \eqref{eq:akshs} together with \eqref{est:Hoelderfourier1}  
	and the equality $A_{k,0}=k (\gamma_N^{(k)}-P^{(k)})$ imply \eqref{est:stongtrace2}.
	
	\underline{Step 3:}
	We have 
	\begin{equation*}
	\tr(A_{k,\theta s})= k (\langle \Psi, S_{1,\theta s} \Psi\rangle-\langle \vphi, S_{1,\theta s} \vphi \rangle)).
	\end{equation*}
	Using the decomposition $1=p_1+q_1$, we obtain
	\begin{equation*}
	\tr(A_{k,\theta s})= k(\langle \Psi, p_1 S_{1,\theta s} p_1 \Psi\rangle-\langle \vphi, S_{1,\theta s} \vphi \rangle +\langle \Psi, q_1 S_{1,\theta s} p_1 \Psi\rangle + \langle \Psi,  S_{1,\theta s} q_1 \Psi\rangle).
	\end{equation*}
	Then observing
	$$\langle \Psi, p_1 S_{1,\theta s} p_1 \Psi\rangle= \langle \vphi, S_{1,\theta s} \vphi \rangle \langle \Psi, p_1 \Psi\rangle
	\leq \langle \vphi, S_{1,\theta s} \vphi \rangle,$$
	it follows that
	\begin{equation*}
	\tr(A_{k,\theta s}) \leq k(\langle \Psi, q_1 S_{1,\theta s} p_1 \Psi\rangle + \langle \Psi,  S_{1,\theta s} q_1 \Psi\rangle).
	\end{equation*}
	Using the decomposition $S_{1,\theta s}=S_{1,\frac{s}{2}} S_{1,\theta s-\frac{s}{2}}$ and  Cauchy-Schwarz, we arrive at the upper bound
	\begin{equation*}
	\tr(A_{k,\theta s}) \leq k(\|S_{1,\frac{s}2}\Psi\|_2+\|S_{1,\frac{s}2}\vphi\|_2)\|S_{1,\theta s-\frac{s}{2}}q_1\Psi\|.
	\end{equation*}
	If $\theta \in(\frac{1}{2},1)$, again applying Hölder's inequality in Fourier space yields
	\begin{align*}\|S_{1, \theta s-\frac{s}{2}} q_1 \Psi\|\;
	&\leq\; \|S_{1,\frac{s}{2}} q_1 \Psi\|^{2 \theta-1} \| q_1 \Psi\|^{2-2\theta}=
	\|S_{1,\frac{s}{2}} q_1 \Psi\|^{2 \theta-1} a_N^{1-\theta}\\
	&\leq\;(\|S_{1,\frac{s}{2}} \vphi\|+\|S_{1,\frac{s}{2}} \Psi\|)^{2\theta-1}a_N^{1-\theta}
	\end{align*}
	If	$\theta \leq \frac{1}{2}$, then trivially we have the non-sharp estimate
	$$\|S_{1, \theta s-\frac{s}{2}} q_1 \Psi\| \leq \|q_1 \Psi\|=a_N^\frac{1}{2}.$$
	The last three estimates imply \eqref{est:stongtrace3} and this completes the proof of Theorem \ref{thm:main1}. 
\end{proof}

\paragraph{Proof that Theorem \ref{thm:main1} implies  part (i) of Theorem \ref{thm:main}}
As pointed out in Remark \ref{rem:intuition}, it is enough to show that there exists $C$ independent of $N,t$ such that 
$$C_{\Psi_{N,t}, \vphi_t, \theta,1} \leq C, \quad \forall N \in \N, t>0.$$
In view of inequality \eqref{est:H1}, this reduces to showing that there exists $C=C(\vphi_0)$ with 
\begin{equation}\label{est:onekinetic}
\langle \Psi_{N,t}, -\Delta_{x_1} \Psi_{N,t} \rangle \leq C, \quad  \forall N \in \N, t>0.
\end{equation}
To see this, note that due to the bosonic symmetry of $\Psi_{N,t}$ we have 
\begin{equation}\label{kinsym}
\langle \Psi_{N,t}, -\Delta_{x_1} \Psi_{N,t} \rangle = \frac{1}{N} \sum_{j=1}^N \langle \Psi_{N,t}, -\Delta_{x_j} \Psi_{N,t} \rangle, \quad \forall N \in \N, t>0.
\end{equation}
Let 
$$v_{ij}(x_1,\dots,x_N):=v(x_i-x_j),$$
where $v$ was defined in \eqref{def:v}. 
Using the inequality $\frac{|\lambda|}{|x_i-x_j|} \leq \frac{1}{4 } \frac{1}{|x_i-x_j|^2} + \lambda^2$ together with Hardy's inequality  it follows that
$$|v_{ij}| \leq \frac{1}{2}(-\Delta_{x_i}-\Delta_{x_j})+\lambda^2,$$
and therefore
$$\left|\frac{1}{N-1}\sum_{i<j} v_{ij}\right| \leq \frac{1}{2} \sum_{j=1}^{N} (-\Delta_{x_j})+ \frac{N \lambda^2}{2}.$$
This gives
$H_N \geq \frac{1}{2} \sum_{j=1}^{N} (-\Delta_{x_j})- \frac{N \lambda^2}{2},$
which together with \eqref{kinsym} implies
$$\langle \Psi_{N,t}, -\Delta_{x_1} \Psi_{N,t} \rangle \leq \frac{2}{N}\langle \Psi_{N,t}, H_N \Psi_{N,t} \rangle + \lambda^2\quad \forall N \in \N, t>0.$$ 
Since moreover $\langle \Psi_{N,t}, H_N \Psi_{N,t} \rangle=\langle \Psi_{N,0}, H_N \Psi_{N,0} \rangle= N \cE(\phi_0)$,  \eqref{est:onekinetic} follows. This concludes the proof of part (i) of Theorem \ref{thm:main}.
\subsection{Proof of part (ii) of Theorem \ref{thm:main}}

	Let $t>0$ be any fixed time. As in step 1 in the proof of Theorem \ref{thm:main1}, we have
	\begin{equation}\label{est:HSH1}
	\tr|S_k^\frac12(\gamma_{N,t}^{(k)}-P^{(k)}_t)S_k^\frac12|
	\leq 2 \|S_k^\frac12(\gamma_{N,t}^{(k)}-P^{(k)}_t)S_k^\frac12\|_{HS}+\tr(S_k^\frac12(\gamma_{N,t}^{(k)}-P^{(k)}_t)S_k^\frac12).
	\end{equation}
	Next we prove
	\begin{equation}\label{trgegen0}
	\tr(S_k^\frac12(\gamma_{N,t}^{(k)}-P^{(k)}_t)S_k^\frac12) \rightarrow 0 \mbox{\ as\ }N\to\infty.
	\end{equation}
	By symmetry of $\Psi_{N,t}$, we obtain
	
	\begin{equation}\label{dec:tr}
	\tr(S_k^\frac12(\gamma_{N,t}^{(k)}-P^{(k)}_t)S_k^\frac12)=k (\langle \Psi_{N,t}, (-\Delta_{x_1}) \Psi_{N,t}\rangle-\langle \vphi_{t}, -\Delta \vphi_{t}\rangle ).
	\end{equation}
	The energy conservation and the initial condition $\Psi_{N,0}=\vphi_0^{\otimes N}$ yield 
	\begin{equation}\label{eq:econs}
	\frac{1}{N} \langle \Psi_{N,t}, H_N \Psi_{N,t} \rangle=\cE(\vphi_0)=\cE(\vphi_t),
	\end{equation}
		and therefore from the symmetry with respect to the particle coordinates it follows that
		\begin{equation}\label{tracepart}
		\langle \Psi_{N,t}, - \Delta_{x_1} \Psi_{N,t}\rangle-\langle \vphi_{t}, -\Delta \vphi_{t}\rangle=- \frac{1}{2}\left(\langle \Psi_{N,t}, v_{12} \Psi_{N,t} \rangle- \langle \vphi_{t}^{\otimes 2}, v_{12} \vphi_{t}^{\otimes 2} \rangle\right).
		\end{equation}
From Kato's inequality $\frac{1}{|x|} \leq \frac{\pi}{2}(-\Delta)^{\frac12}$, see \cite{Her}, it follows that 
$S_2^{-\frac14}v_{12}S_2^{-\frac14}$ can be extended to a bounded operator on $L^2(R^3)^{\otimes_S2}$. 
Part (i) of Corollary \ref{cor:application} follows from part (i) of Theorem \ref{thm:main}, which we already proved. Hence we can apply it to $\cA=v_{12}$ to obtain 
	$$\lim_{N \to \infty}	\langle \Psi_{N,t}, v_{12} \Psi_{N,t} \rangle= \langle \vphi_{t}^{\otimes 2}, v_{12} \vphi_{t}^{\otimes 2} \rangle \quad \forall t>0,$$
	which together with \eqref{dec:tr} and \eqref{tracepart} gives \eqref{trgegen0}.

   It remains to show
	\begin{equation}\label{HSgegen0}
	\lim_{N\to\infty}\|S_k^\frac12(\gamma_{N,t}^{(k)}-P^{(k)}_t)S_k^\frac12\|_{HS} =0. 
	\end{equation}
	We observe
	$$ \|S_k^\frac12 \gamma_{N,t}^{(k)}S_k^\frac12\|_{HS} \leq \tr(S_k^\frac12 \gamma_{N,t}^{(k)}S_k^\frac12)$$
	and 
	$$ \|S_k^\frac12 P^{(k)}_t S_k^\frac12\|_{HS} = \tr(S_k^\frac12 P^{(k)}_t S_k^\frac12).$$
	Thus, \eqref{trgegen0} implies that
	\begin{equation}\label{est:limsup}
	\limsup_{N\to\infty} \|S_k^\frac12 \gamma_{N,t}^{(k)}S_k^\frac12\|_{HS} \leq \|S_k^\frac12 P^{(k)}_t S_k^\frac12\|_{HS}.
	\end{equation}
	In particular, $(S_k^{1/2} \gamma_{N,t}^{(k)}S_k^{1/2})_{N\in\N}$ is bounded in the Hilbert-Schmidt norm and so it has a weakly convergent subsequence, also denoted by $(S_k^{1/2} \gamma_{N,t}^{(k)}S_k^{1/2})_{N\in\N}$, weakly converging to some Hilbert-Schmidt operator $K$. Consequently, $(\gamma_{N,t}^{(k)})_{N\in\N}$ weakly converges to $S_k^{-{1/2}} K S_k^{-{1/2}}$. As we mentioned in the introduction, it is by now established and well-known that we have
	\begin{equation*}
	\lim_{N\to\infty} \|\gamma_{N,t}^{(k)}-P^{(k)}_t\|_{HS} \,=\,0,
	\end{equation*}
	see for example Theorem \ref{thm:CLS},
	 and thus we obtain $K=S_k^{1/2} P^{(k)}_t S_k^{1/2}$. Using now \eqref{est:limsup}, we arrive at
	$$S_k^{\frac12} \gamma_{N,t}^{(k)}S_k^{\frac12}\stackrel{\|\cdot\|_{HS}}{\longrightarrow}S_k^{\frac12} P^{(k)}_t S_k^{\frac12}\quad\mbox{\ as\ }N\to\infty.$$
	Thus we have shown that any subsequence of $(S_k^{1/2} \gamma_{N,t}^{(k)}S_k^{1/2})_{N\in\N}$ has a further subsequence converging to $S_k^{1/2} P^{(k)}_t S_k^{1/2}$ in the Hilbert-Schmidt topology, which proves \eqref{HSgegen0}. \eqref{est:HSH1}, \eqref{trgegen0} and \eqref{HSgegen0} imply part (ii) of Theorem \ref{thm:main}.

\end{document}